\documentclass[12pt]{article}

\usepackage{amssymb}
\usepackage{graphicx}
\usepackage{subfigure}

\begin{document}

\begin{titlepage}

\begin{flushright}
arXiv:1104.5673
\end{flushright}
\vskip 2.5cm

\begin{center}
{\Large \bf Bounds on Parity Violation\\
In the Cosmological Redshift}
\end{center}

\vspace{1ex}

\begin{center}
{\large Brett Altschul\footnote{{\tt baltschu@physics.sc.edu}}}

\vspace{5mm}
{\sl Department of Physics and Astronomy} \\
{\sl University of South Carolina} \\
{\sl Columbia, SC 29208} \\

\vspace{10mm}
{\large Matthew Mewes}

\vspace{5mm}
{\sl Department of Physics and Astronomy} \\
{\sl Swarthmore College} \\
{\sl Swarthmore, PA 19081} \\
\end{center}

\vspace{2.5ex}

\medskip

\centerline {\bf Abstract}

\bigskip

Parity (P) violation in interactions between the spacetime metric and other
fields would be a sure sign of new physics. We examine the
possibility of P violation in the cosmological redshift. If right- and left-circularly
polarized photons experience the redshift differently, the radiation from distant
sources would tend to be depolarized, since the polarization states would accumulate
different phases during propagation. The resulting birefringence has an unusual
signature---depending on $z^{2}$---quite unlike what is seen in other theories,
including those with violation of local boost invariance.
The observed broad-spectrum polarization of
$\gamma$-ray bursts constrains the fractional difference between the right- and
left-handed redshifts at the $6\times10^{-37}$ level.

\bigskip

\end{titlepage}

\newpage

\section{Introduction}

The search for new fundamental physics, especially a quantum theory of gravitation,
is an area of fundamental importance. However, finding direct evidence for quantum
gravity is expected to be a very difficult problem, because the Planck scale $M_{P}$
is so large compared with the energy scales accessible in particle physics.
An alternative way of approaching this problem is to search for evidence of new
physics at low energies, by looking for phenomena that cannot occur within
the physical theories we already understand---the standard model and general
relativity. Such phenomena include violations of charge conservation, Lorentz
symmetry, CPT, and the spin-statistics relationship; all these possibilities have been
searched for and not seen. However, if any of these exotic phenomena were observed in
the laboratory, that would be a discovery of paramount importance and a powerful clue
about the nature of new fundamental physics. In this paper, we shall examine another
such exotic possibility---a manifestation of parity (P) violation in
the cosmological redshift.

There has recently been a substantial amount of work looking at the possibility of
P violation in gravity---a phenomenon that is small in the standard
model~\cite{ref-chu}. Many analyses have focused on Chern-Simons
gravity~\cite{ref-jackiw-pi}, in which the Einstein-Hilbert action is supplemented
with a P-violating term that is second order in the curvature. Such a theory could
be constrained with measurements of gravitomagnetic
effects~\cite{ref-alexander1,ref-smith1,ref-yunes1}.
Other searches have looked for evidence of gravitational P violation in
conjunction with Lorentz symmetry
breaking~\cite{ref-battat,ref-muller5,ref-bailey2,ref-panjwani}.
In phenomenalistic models (not necessarily
described by local field theories) in which right- and
left-polarized gravitational waves couple differently, it was also
found that there would be P-violating imprints on the cosmic microwave
background~\cite{ref-lue,ref-contaldi}.

It is also possible to study P violation in
phenomena that are not related to the gravitational dynamics but rather to the
metrical structure of spacetime.
The redshift measures the expansion of the universe, which is the single most
important phenomenon in cosmology. We shall examine the possibility that the redshift
affects right- and left-circularly polarized radiation differently. (Such a difference
obviously breaks P invariance and potentially other symmetries as well.)
If the redshift is an effect purely of geometry, then it should be the same for
all electromagnetic radiation; the wavelength is simply stretched out by a
constant factor, while waves' local propagation speeds remain uniform. However, the
redshift can also be viewed as the result of continual and coherent
rescattering of electromagnetic radiation by a time-dependent
metric field $g^{\mu\nu}$. If there is P violation in the
photon-metric interaction, it could manifest itself as a difference in
redshifts for right- and left-circularly polarized radiation. Like most effects that
treat the two polarization states dissimilarly, P violation in the redshift can
lead to birefringence, which may be studied using spectropolarimetry.

There are two ways that a P-violating redshift could depolarize the
radiation that reaches us across cosmological distances.
The first way is through the redshift $z=\frac{\lambda_{o}}{\lambda_{e}}-1$
(defined in terms of a wave's wavelength at the times of emission and observation)
itself. If
the left- and right-polarized components of an originally monochromatic wave have
different final wavelengths, the wave cannot be purely linearly polarized.
However, this turns out to provide a much weaker test of P invariance than another
effect. If the wavelengths of the two polarizations differ during propagation, they
will accumulate different phases by the time they are detected. The resulting
frequency-dependent difference between the final phases for the two circular
components could easily destroy any linear polarization in a broad-spectrum
source.

Violation of P (and CP) symmetry in the redshift is a far-reaching effect and would
require
substantial modification of general relativity. (A non-metric theory of gravitation
might be required to accommodate such an effect.) However, we shall not specify the
form of the underlying theory responsible for the P violation. Instead, we shall
attempt to describe largely model-independent bounds on the possibility.

It is important that the effect we are considering (like that discussed
in~\cite{ref-contaldi}) cannot occur in the
framework of local field theory, and so the existence of the effect would signal the
presence of new physics of a completely novel character.
In a local field theory, photon propagation is governed by a gauge-invariant
Lagrange density ${\cal L}(g,A)$, which is bilinear in the electromagnetic field $A$
and also depends on the (slowly varying) metric $g$. The equivalent problem of a
bilinear electromagnetic Lagrange density containing arbitrarily many derivatives was
considered in~\cite{ref-kost23}, and it was found that no local
theory could produce energy-independent differences between the propagation of right-
and left-circularly polarized waves.

Having different phase speeds $c_{\pm}$ for right- and left-polarized waves produces
an effect qualitatively similar to P violation in the redshift, but the signatures of
the two phenomena are quite distinct.
There are already extremely tight constraints on the kind of birefringence that would
be caused by the two polarizations having different phase speeds.
Such a variation in the speed of light violates local Lorentz boost invariance, and
it can be ruled out at approximately the $10^{-37}$
level~\cite{ref-kost23,ref-kost21}. The cited analysis considered only a basis of
linear polarization states (although the circular state analysis and results are
similar), since an energy-independent difference in phase speeds for the two circular polarization states does not occur in local field theory.
However, energy-dependent speed
differences between states of circular polarization are possible and have been
constrained~\cite{ref-carroll1,ref-gleiser,ref-jacobson2}.

\section{Parity Violation in the Redshift}

The effective distance that a photon travels between emission and detection depends
on the redshift of the source and the expansion history of the universe. However, the
situation simplifies for small redshifts. At lowest order, this distance is then
simply $L_{0}\approx z/H_{0}$.
If right- and left-circularly polarized waves take different
times to cross a constant distance $L_{0}$,
this is precisely the same as their having different
propagation speeds. Despite the presence of $z$ in the formula for $L$,
the actual redshift effect is not involved here; instead, we have simply exploited
Hubble's Law, which states that $z$ and $L_{0}$ are proportional.
The effect we are looking to constrain is different; it entails the two polarizations
having different observed redshifts, because their wavelengths expand
at rates slightly different from the overall expansion rate of the universe.

The redshift $z(t)$ relates the energy of a given polarization of a wave at
different times in its history. For a wave emitted at time $t_{e}$ with energy
$k_{e}$ (the same for both polarizations), its energy at a later time $t'$ is related
by
\begin{equation}
\frac{k(t')}{1+z(t')}=\frac{k_{e}}{1+z(t_{e})};
\end{equation}
under ordinary circumstances, $z(t_{e})$ is simply the redshift $z$ of the source.
However, we are interested in what happens if
P violation manifests itself through the right and left polarizations having
systematically different $z$ values, $z_{\pm}$. In that case, the two polarizations'
energies will vary differently with time, so that
\begin{equation}
\delta k(t')\equiv k_{+}(t')-k_{-}(t')=\left[\frac{1+z_{+}(t')}{1+z_{+}(t_{e})}-
\frac{1+z_{-}(t')}{1+z_{-}(t_{e})}\right]k_{e}.
\end{equation}
In terms of the average redshift $z(t)$ and $\delta z=z_{+}-z_{-}$, we have
$z_{\pm}=\left(1\pm\frac{1}{2}\frac{\delta z}{z}\right)$.
We shall assume $\frac{\delta z}{z}$ is small and constant. (If it is not constant,
our final observable will depend on an appropriately weighted average of the
$\frac{\delta z}{z}$ values relevant to the propagation period.) It follows that
\begin{equation}
\delta k\approx\left(\frac{\delta z}{z}\right)\frac{z'-z}{(1+z)^{2}}k_{e}
\approx\left(\frac{\delta z}{z}\right)\frac{z'-z}{(1+z)}k_{o},
\end{equation}
where $z'=z(t')$, $z=z(t_{e})$ is the redshift of the source, and $k_{o}$ is the
energy at the time of observation. The accumulated phase difference between the
polarizations at the time of observation is
\begin{equation}
\label{eq-deltaphi}
\delta\phi=\int_{t_{e}}^{t_{o}}dt'\,\delta k(t')=\left(\frac{\delta z}{z}\right)
\frac{k_{o}}{1+z}\int_{0}^{z}dz'\,\frac{z'-z}{(1+z')H(z')},
\end{equation}
in terms of the Hubble expansion parameter $H(z')$.
The plane of polarization for a linearly
polarized photon is rotated through an energy-dependent angle
$\delta\psi=\frac{1}{2}\delta\phi$.

For our calculations, we shall use the full formula (\ref{eq-deltaphi}) for
$\delta\phi$. However, the behavior of $\delta\phi$ for $z\ll 1$ is particularly
simple and has a straightforward interpretation.
For small $z$, the common phase between the two polarizations is $k_{o}L_{0}$,
where $L_{0}$ is the instantaneous distance at the time of emission. The redshift
adds to the total travel distance during the propagation time.
Since $z\ll 1$, the rate at which extra distance is added to
the path is $H_{0}$ times the distance remaining between a traveling
wave's position and the Earth. Integrated over the travel time of approximately
$L_{0}$, this gives a formula for the distance that is corrected by the expansion of
the universe:
\begin{equation}
\label{eq-L}
L\approx L_{0}+\frac{1}{2}H_{0}L_{0}^{2}\approx L_{0}+\frac{1}{2}zL_{0}.
\end{equation}
Since $z$ differs between the two polarization, the phase associated with the added
propagation distance $\frac{1}{2}zL_{0}$ differs between the
two polarization distances by $-\left(\frac{\delta z}{z}\right)k_{o}$ times
$\frac{1}{2}H_{0}L_{0}^{2}$.
The phase difference is therefore
\begin{equation}
\delta\phi\approx
-\frac{1}{2}\left(\frac{\delta z}{z}\right)k_{o}\frac{z^{2}}{H_{0}}.
\end{equation}
For small $z$, $\delta\phi$ is
proportional not to the distance to the source, but to the distance squared;
this is a signature unlike any that would have been seen in theories with a
polarization-dependent speed of light. Of course, the characteristic behavior for
small $z$ can also be derived from a direct expansion of (\ref{eq-deltaphi}).

\section{Constraints From $\gamma$-Ray Bursts}

The sustained
linear polarizations of the $\gamma$-ray bursts GRB 930131, GRB 960924, and GRB
041219a over wide photon energy ranges can be used to rule out a polarization
dependence in the redshift quite strongly. Since the rotation of the polarization
plane is energy dependent, the existence of a consistent polarization direction
across a broad range of energies strongly constrains $\frac{\delta z}{z}$. Such
constraints are possible even when the initial polarization direction is unknown.

GRB 930131 and GRB 960924 had measured polarizations over an observed energy range
of 31--98 keV of
at least 35\% and 50\%, respectively~\cite{ref-willis}, and each source had a
redshift of at least
$z=0.1$. For GRB 041219a, the observed polarization was at least 33\% (at $1\sigma$)
in a 100--350 keV energy band~\cite{ref-mcglynn}, and generally consistent results
were found in higher-energy bands. The estimated redshift for this burst was
$z\approx0.7$; this estimate was made using a known correlation between the spectral
peaks and luminosities of $\gamma$-ray bursts~\cite{ref-yonetoku}.

\begin{figure}
\centering
\includegraphics[scale=0.75]{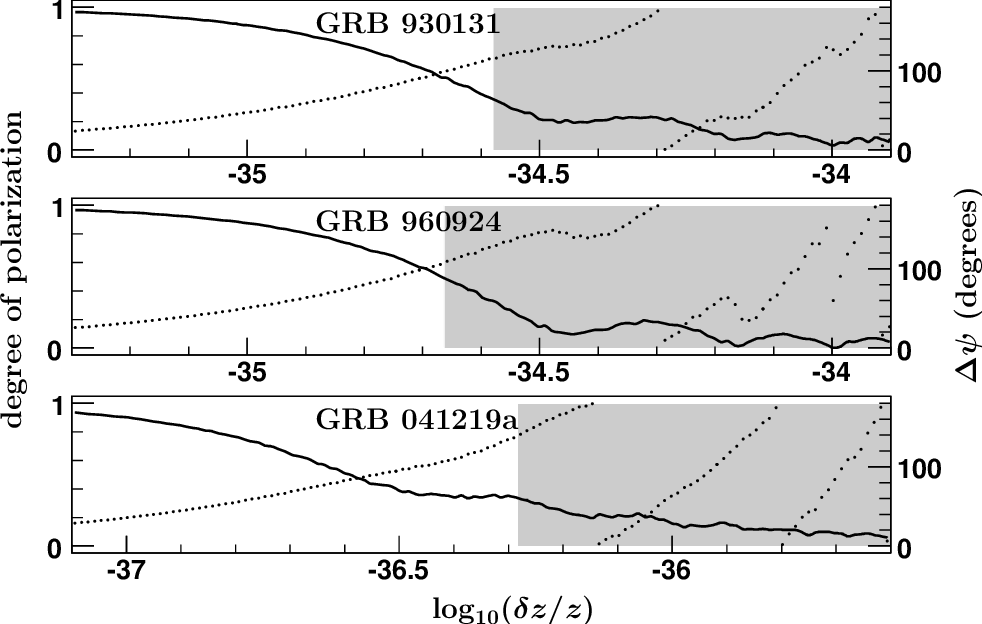}
\caption{The observed degree of polarization $\Pi$ (solid line) and rotated
polarization angle in degrees (dots) for GRB 930131, GRB 960924, and GRB 041219a,
assuming the radiation is 100\% linearly polarized at the source,
as a function of $\log_{10}(\delta z/z)$. The shaded region is excluded by the
bursts' observed levels of polarization.
\label{fig-grb}}
\end{figure}

We used these data to rule out insufficiently small values of $\frac{\delta z}{z}$.
The most conservative assumption we could make about the initial polarization
of the sources was that the radiation began 100\% linearly polarized, before any
birefringence occurred. We have
calculated numerically the polarization that would remain to be observed at the
Earth, smearing the energy-dependent birefringence over the observed energies.
The analysis was performed separately for each source, according to the following
procedure. For each one of a large number of possible $\frac{\delta z}{z}$ values,
we simulated $10^{4}$ photons. For GRB 930131 and GRB 960924, the
photons were distributed across the observational energy range according to a
polynomial fit to the measured fluxes. For GRB 041219a, we used a Band
model~\cite{ref-band} that was known to provide a good fit to the burst's
spectrum~\cite{ref-mcbreen}. We then calculated the total polarization fraction
and the rotation of the polarization angle for each sample. We used the
Hubble parameter for a $\Lambda$CDM cosmology with $H_{0}=72$
km$\cdot$s$^{-1}\cdot$Mpc$^{-1}$, $\Omega_{m}=0.26$ and $\Omega_{\Lambda}=0.74$.
Figure~\ref{fig-grb} shows the results of these calculations---the surviving degree
of polarization (and the rotation of the polarization plane)
as a function of $\frac{\delta z}{z}$, including the values of
$\frac{\delta z}{z}$ that are ruled out by the respective sources' observed 35\%,
50\%, and 33\% polarizations. The decay in the observed polarization with increasing
$\frac{\delta z}{z}$ is pronounced but not strictly monotonic. For small values of
$\frac{\delta z}{z}$, the polarizations at different frequencies diverge, but for
sufficiently large values of the parameter, the polarizations of different parts of
the spectrum may reconverge, when their total rotation angles $\delta\psi$ happen to
differ by a multiple of $\pi$.

The characteristic scale of the resulting bounds is
\begin{equation}
\label{eq-estimate}
\left|\frac{\delta z}{z}\right|\lesssim\frac{4}{\Delta kLz}
\end{equation}
(where $\Delta k$ is the energy range covered by the polarization measurements),
indicating bounds at the $10^{-37}$--$10^{-36}$ level. Taking the polarization and
distance parameters given above,
the detailed calculations rule out a $\left|\frac{\delta z}{z}\right|$ larger than 
approximately $3\times10^{-35}$ using the GRB 930131 or GRB 960924 data and the
somewhat stronger
\begin{equation}
\label{eq-bound1}
\left|\frac{\delta z}{z}\right|<6\times10^{-37}
\end{equation}
with GRB 041219a. The result (\ref{eq-bound1}) is our primary bound on P violation in
the redshift.

\section{Other Effects of $\delta z/z$}

It is worthwhile to examine whether another cause of depolarization
might also be important experimentally. If the right- and left-circular polarizations
from an originally monochromatic wave arrive at the Earth with slightly different
wavelengths, the wave train they form cannot remain coherent over its entire length
$D$. However, we shall see that the resulting depolarization effect is weaker than
the one discussed above by a enormous factor of $L/D$.

An initially linearly polarized monochromatic wave is composed of equal
amounts of right- and left-handed radiation,
$\vec{E}=\frac{E_{0}}{\sqrt{2}}
[\hat{\epsilon}_{+}e^{ik_{e}(x_{3}-t)}+\hat{\epsilon_{-}}e^{ik_{e}(x_{3}-t)}]$. If
the circular polarization states are subject to different redshifts $z_{+}$ and
$z_{-}$, the components will reach the Earth with different wave vectors
$k_{\pm}=k_{e}/(1+z_{\pm})$. If we consider only the difference in frequencies
(rather than the phase shifts discussed above), the redshifted wave has the form
\begin{eqnarray}
\vec{E} & = &
\frac{E_{0}}{\sqrt{2}}\left[\hat{\epsilon}_{+}e^{ik_{+}(x_{3}-t)}+\hat{\epsilon_{-}}
e^{ik_{-}(x_{3}-t)}\right]
\\
\label{eq-Erotating}
& = & E_{0}e^{ik_{o}(x_{3}-t)}\left\{\hat{x}_{1}\cos\left[\frac{k_{\Delta}}{2}(x_{3}
-t)\right]-\hat{x}_{2}\sin\left[\frac{k_{\Delta}}{2}(x_{3}-t)\right]\right\}.
\end{eqnarray}
Here $k_{o}=(k_{+}+k_{-})/2$ and $k_{\Delta}=k_{+}-k_{-}$. At $k_{o}(x_{3}-t)=0$, the
wave maintains its linear polarization along the $x_{1}$-direction. However,
ahead of and behind this wave front, the polarization of the wave train is
rotated. The wave train as a whole will be strongly polarized
only if this rotation is approximately less than $\frac{\pi}{2}$ over the
length $D$ of the wave train.
More precisely, the Stokes parameters for the wave described by (\ref{eq-Erotating}),
over a wave train of length $D$ centered around $x_{3}=t$ are
$\langle s^{2}\rangle=\langle s^{3}\rangle=0$, and
\begin{equation}
\langle s^{1}\rangle=\frac{\sin(k_{\Delta}D/2)}{k_{\Delta}D/2}I_{0}=\pm\Pi I_{0},
\end{equation}
where $I_{0}$ is the intensity, and $\Pi$ is the degree of polarization, which
falls off as $(k_{\Delta}D)^{-1}$.


To determine the wave train length $D$, we must look at how the polarized radiation
is emitted. Synchrotron electrons revolve with frequencies
$\omega_{B}=\frac{eB}{\gamma m}=\frac{\omega_{c}}{\gamma}$. An
orbiting electron radiates up to a cutoff frequency
$\omega_{C}=\frac{3}{2}\gamma^{3}\omega_{B}\sin\alpha$, where
$\gamma$ is the Lorentz factor of the electron and
$\alpha$ is the pitch angle
between the direction of $\vec{B}$ and the particle velocity $\vec{v}$.
An electron of
energy $\gamma m$ emits most of its radiation near $\omega_{C}$, above
which frequency the radiated power falls off rapidly. Conversely,
radiation with frequency $\omega$ comes mostly from electrons with Lorentz
factors $\gamma\sim(\omega/\omega_{c})^{1/2}$.
The radiation from synchrotron electrons is strongly beamed along the
direction of their motion. The beam has an angular width of
$\Delta\theta\sim\frac{2}{\gamma}$. During each revolution, the charge's 
direction
of motion sweeps through a range of angles $2\pi\csc\alpha$, and the narrow beam of
radiation flashes along the line of sight for a brief time
$D_{0}\approx\frac{2}{\gamma\omega_{B}}\csc\alpha=\frac{2m}{eB}\csc\alpha$.
The range of angles into which the radiation is emitted depends
relatively little on $\omega$ when $\omega\lesssim\omega_{C}$, and the length of
the wave train is essentially the pulse duration $D_{0}$ for all relevant
emission frequencies.

The observation of a wave with linear polarization $\Pi$ on the Earth requires that
$k_{\Delta}D<\frac{2}{\Pi}$. Since $k_{\Delta}=k_{o}(\delta z)$, this produces a
bound on $\frac{\delta z}{z}$ that is weaker than (\ref{eq-estimate}) by factor of
$\sim L/D$. For an optimal source (such as the quasar 3C 273, located at $z=0.158$,
where there is clear evidence of polarized synchrotron emission extending from the
radio range up into the optical~\cite{ref-roser3}), $\frac{\delta z}{z}$ would
need to be at the $10^{-10}$ level or larger for its effects to be detectable in
this way.

Inverse Compton (IC) upscattering can increase the energies of polarized synchrotron
photons by as much as $4\gamma^{2}$, but this also tends to decrease the wave
train length $D_{0}$. In any case, there are no cosmic electrons energetic enough
to produce IC photons for which the depolarization that follows
solely from a difference in observed wavelengths would be rapid enough to
generate bounds competitive with (\ref{eq-estimate}) and (\ref{eq-bound1}).

The existence of differing final energies $k_{\pm}$ would also lead to a splitting
of the spectral lines seen in redshifted sources. However, direct comparisons of
the wavelengths of right- and left-polarized radiation would give constraints many
orders of magnitude weaker than (\ref{eq-bound1}).

\section{Conclusion}

Just as understanding P violation is very important to our understanding of
particle physics, any P violation involving couplings to the spacetime metric
would be extremely
interesting. We have derived a strong bound on any possible P violation in the
cosmological redshift, confirming the basically geometric character of this
phenomenon. Although the bound (\ref{eq-bound1}) is already very strong, significant
improvement will still be possible, as polarization measurements from higher redshift
$\gamma$-rays busts become available. Bursts have been observed with redshifts
substantially greater than 1 and energies well above 1 MeV, and their true degrees
of polarization may be close to 100\%. This indicates that the possible improvement
could be substantial, although the necessary measurements will naturally be
challenging.

\end{document}